\documentstyle[12pt]{article}
\def \F{\phi}

\def \T{\theta}
\def \P{\psi}
\def \D{\delta}

\def \G{\Gamma}
\def \g{\gamma}

\def\NP{{\it Nucl. Phys.\ }}

\def\PL{{\it Phys. Lett.\ }}

\def\PRL{{\it Phys. Rev. Lett.\ }}

\def\e{\epsilon}

\def\a{\alpha}
\def\l{\lambda}

\def\s{\sigma}

\def\half{{1\over 2}}

\def\be{\begin{equation}}
\def\eq{\end{equation}}
\def\Tr{{\rm Tr}}
\def\cA{{\cal A}}

\begin{document}

\begin{flushright}
OUTP-9419P\\
cond-mat/9409011\\
\end{flushright}
\vspace{20mm}
\begin{center}
{\LARGE Polymer Collapse on Fluctuating Random Surfaces}\\
\vspace{30mm}
{\bf Simon Dalley}\\
\vspace{5mm}
{\em Department of Physics, Theoretical Physics\\
Oxford University, Oxford OX1 3NP, U.K.}\\
\end{center}
\vspace{30mm}
\abstract
The conformations of interacting linear polymers  on a dynamical planar random
lattice are studied using a random two-matrix model. An exact
expression for the partition function of self-avoiding chains
subject to attractive contact interactions of relative strength $1/c$,
$0<c<1$, is found as a function of chain length $L$ and $c$.
The number of configurations $L^a {\rm e}^{bL}$ as $L\to \infty$ is
determined, showing that a chain undergoes a third-order collapse
transition at $c=\sqrt{2} -1$; the universal scaling laws are found.

\newpage
\section{Introduction}
\baselineskip .3in
A long polystyrene chain molecule in a solvent
undergoes
a coil--to--globule transition \cite{coil}
 at a finite temperature $T = \Theta$ due to
attractive
polymer--polymer interactions. Many approximate theoretical techniques
have been developed to describe this and other processes of
macromolecular folding \cite{dg}. In this letter
 exact results for this problem of polymers, modeled by self-avoiding
chains
 with contact interactions,
will be given on
the ensemble of
two-dimensional simplicial lattices with random fluctuations of the
local
intrinsic curvature which appear in the study of
two-dimensional quantum gravity \cite{jan,rest}. This ensemble represents an
annealed average over a certain class of fractals. The advantage is
that statistical mechanics problems are often
exactly solvable in this case
using random matrix models \cite{rest}, which yield
expressions for the thermodynamic and some of the geometrical quantities of
interest.
Moreover it is intuitively plausible, and
found
to be the case in all of the many solved examples, that the phase
structure of such statistical systems
is the same as on a fixed lattice
but with weaker transitions when one averages over
fluctuating
lattices.
Although a simple structureless polymer will be studied here,
the same techniques could be extended to more complicated
polymers types.

A two-dimensional simplicial lattice is obtained by gluing, pairwise
along
edges, $A$ squares each of edge length $\D$,
so as to form a surface with spherical topology say.
A  polymer chain of $L$ links walks isotropically
on the edges of squares, cannot cross itself
(excluded
volume), but can occupy a given edge any number of times (figure 1);
the latter represent contacts. The partition function is a sum over
all possible simplicial lattices and configurations of a polymer on
such lattices with a weight for contacts.
The $\Theta$-temperature is found exactly in this model and represents a
third-order collapse transition. One is therefore led to predict that
on a fixed two-dimensional lattice (e.g. honeycomb),
with the rules of contact given in Fig.1, the transition will
be first or second order.\footnote{Some previous exact results \cite{dup}
on fixed lattices have used a different definition
of contact, forbidding entirely configurations `a' and `b' of Figure
1;
a contact occurs when two non-adjacent
parts of the polymer come within one lattice
spacing. This could make a difference in the collapsed regime
since the polymer can be much more compact if multiple occupation of
an edge is allowed (type `a' in Figure 1). Such multiple occupation
necessarily occurs, for example, in the high-temperature expansion of
the lattice $O(n)$ model with action
$S.S$}
 The
universal
exponent $a$, defined by the number of configurations as $L \to
\infty$,
$L^a {\rm e}^{bL}$, changes from $a=-5/2$ at $T > \Theta$ to $a=-3/2$ at
$T = \Theta$. For $T < \Theta$
it was not possible to determine $a$.
The universal behaviour around
$T=\Theta$ is described in terms of a reduced temperature $T-\Theta \sim
L^{-1/6}$.

\section{Matrix Model.}
The self-avoiding walk on fluctuating random surfaces was described
using a one-matrix model by Duplantier and Kostov \cite{kostov}. In order to
include an attractive interaction whenever contacts occur
consider a two-matrix model with hermitian
$N$x$N$ matrices $\F$ and $\P$,
\be
Z  = \int {\cal D}\F {\cal D}\P {\rm exp}\left( N\Tr \left[ \half \F^2
+ \half \P^2 -c\F \P -\frac{g(1-c^2)^2}{4}\F^4 +(1-c^2)^L \G
\P^{2L}\right] \right) \ .  \label{matmod}
\eq
As usual, the Feynman diagram representation of the expansion in $g$
generates graphs dual to simplicial lattices and as $N \to \infty$ those
of genus zero dominate. $\G$ is a chemical potential conjugate to the
number of polymers $n$, each of length $L$, i.e the surfaces are built
from
$\F^4$ vertices dual to squares while the polymers from $\P$ matrices.
The last term in in the action $\Tr V(\F , \P )$
represents a hole of length $2L$ in the surface
with one point on the boundary marked; one can visualize the polymer
by sewing up the hole starting at the marked point. Each
possible
hole and each possible location of the marked point give a different
polymer configuration. The polymer is connected into the surfaces by
$\F \P$ so that $c$ represents a polymer-lattice Boltzmann weight,
while
the polymer-polymer (contact) weight $\P^2$ has been normalised to one.
Thus $0<c<1$ sets the temperature scale, with $c=1$ being infinite temperature.
\be
Z_{N \to \infty} = \sum_{w,n,A} c^w \G^n g^A H(A,w,n,L) \nonumber
\eq
where the sum is over genus zero surfaces and $H$ is the number of
configurations with $n$ polymers, each of length $L$ with $w$
polymer-contacts, on all surfaces of area $A$.
Since $H$ grows as $\sim
{\rm const}^A$ there exists a critical $g=g_c$ at which large area
surfaces become significant and a universal continuum limit can be
established \cite{rest}.

The matrix model (\ref{matmod}) can be solved in the large $N$ limit
by
the method of orthogonal polynomials \cite{mehta}. The polynomials $P_{i}(\F)
=\F^i + \cdots$ , $Q_{j} (\P) = \P^j + \cdots $ are such that
\be
\int_{-\infty}^{+\infty} d\F d\P {\rm e}^{-V(\F , \P )}
P_i (\F) Q_{j} (\P) = h_i \D_{ij} \label{ortho}
\eq
and the free energy is $ -\log{Z} =  {\rm const} - \sum_{i=1}^{N-1}
(N-n) \log{f_i}$,
where $f_i = h_{i}/h_{i-1}$ and the constant is $g$-independent.
$P$ and $Q$ satisfy recursion relations
\begin{eqnarray}
\F P_{i} (\F) & = & P_{i+1} + r_i P_{i-1} + s_i P_{i-3} + \cdots \\
\P Q_{j} (\P) & = & Q_{j+1} + q_j Q_{j-1} + t_j Q_{j-3} \label{recur}
\end{eqnarray}
That for $P$ terminates after $L+1$ terms. From identities such as
\be
\int d\F d\P {\rm e}^{-V(\F,\P)} P_{i-1}' Q_i  =  0 \nonumber
\eq
one can find a closed set of equations for $r,s,q,t,$ and $f$ which
can be solved for $f$.
In the large $N$ limit one can define a continuous variable $x=i/N$
and functions $f_i = Nf(x), r_i = Nr(x), $ etc. and these equations are
explicitly:
\begin{eqnarray}
cr & = & f-\G \sqrt{f} \T [i,i-1] \label{one}\\
cq & = & f -3g(1-c^2)^{2}fr \label{two}\\
ct & = & -g(1-c^2)^{2}f^3 \label{three}\\
x & = & -cf +q -\G \sqrt{f} \T [i-1,i]\label{five}\ ,
\end{eqnarray}
where
\be
\T [i,j] = 2L(1-c^2)^L \int d\F d\P {\rm e}^{-V(\F,\P)}
\frac{P_i}{\sqrt{h_i}} \P^{2L-1} \frac{Q_j}{\sqrt{h_j}}\ .
\eq
Substituting eqs. (\ref{one}) and (\ref{two}) into eq.
(\ref{five})
gives an equation (the `string equation' in quantum gravity) for $f$.
$\T$ will be evaluated shortly for large $N$.

The critical  point for continuum surfaces is identified by setting $\G=0$
(no polymers) and finding the singularity in the solution for $f(1)$
of the quadratic equation (\ref{five}) as $g$ is varied,
yielding $g_c ={1 \over 12}$ and $f_c = 2c/(1-c^2)$.
Defining $g=g_c (1-\mu \D^2)$, where $\D$ is the lattice spacing,
$\mu$
is the renormalised 2D `cosmological constant' conjugate to
renormalised
surface area $\cA = A\D^2$. In the limit $\D \to 0$, $\mu$ couples to
universal continuum surfaces ($\cA$ finite) and surfaces finite in
lattice
units ($\cA =0$). The partition function will have a part regular in
$\mu$, usually representing the latter surfaces, and a non-analytic
part representing the former.
It is appropriate to define scaling variables $x= 1- z\D^2$ and $ f(x)
= f_c
( 1- u(z) \D)$
in which case, in the limit $\D \to 0$,
\be u^2 = \mu + z \label{string}
\eq
from eq.(\ref{five}) at $\G =0$ and
\be
- \log{Z} = N^2 \D^5 \int_{0}^{\infty} dz \ z u \ + {\rm regular}
\eq

\section{Polymer  Collapse.}
First the case  of a single polymer
will be addressed, when  eq.(\ref{string}) is appropriate
and one studies the expectation value of $\Tr(\P^{2L})$ in this
theory, i.e. $Z$ expanded to first order in $\G$, $Z= Z_0 + \G Z_1 +
\cdots$,
\begin{eqnarray}
Z_1 & = & \int
{\cal D}\F {\cal D} \P (1-c^2)^L \Tr(\P^{2L}) {\rm e}^{ -N\Tr (V(\G
=0))} \nonumber\\
& =& (1-c^2)^L \sum_{i=1}^{N} \int d\F d\P {\rm e}^{-V(\G=0)}
{P_i \over \sqrt{h_i}} \P^{2L} {Q_i \over \sqrt{h_i}} \ .
\end{eqnarray}
Thus, if $Z_1 = \sum_{i=1}^{N} Z_{1}'$, from
eqs.(\ref{recur})(\ref{ortho})(\ref{two}) and (\ref{three})
one finds
\begin{eqnarray}
Z_{1}'(L,c,u,\mu) & = & (1-c^2)^L (2L)! f^L
\sum_{p=0}^{[L/2]} {( c -3gf)^{L-2p}
(-gf)^{p} c^{3p -2L} \over (L-2p)! (p+L)! p!}
\label{sum}\\
{gf \over c} & =& {1-c^2 \over 6} (1- u\D + O(\D^2)) \nonumber
\end{eqnarray}
($[L/2]$ indicates
greatest integer $\leq L/2$).
In order that the polymer be an extended object on the surfaces of
finite
$\cA$ one must scale its length as $L = l / \D$ in which case
\be
Z_{1}' = G(L,c) {\rm e}^{-\l (c) ul + O(\D)} + O({\rm
const}^{-O(\D^{-1})}) \label{asym}
\eq
where
$G \propto L^a {\rm e}^{b(c) L}$. Despite many attempts the author has
not managed to {\em derive} the asymptotic form (\ref{asym}) for all
$c$.
It is generally expected however that the number of configurations has
the
form of $G$ as $L \to \infty$, which numerical evaluation of
(\ref{sum}) confirms. Figure 2 plots $\l (c)$ obtained from
\be
ul\l = \log{\left[ {Z_{1}'(L,c,0) \over Z_{1}'(L,c,ul)} \right]}
\label{derive}
\eq
for large but finite $L$ --- corrections to the right hand side
above are $O(\D)$.

$G = \lim_{L \to \infty} Z_{1}'(L,c,0)$
is a kind of wavefunction renormalisation which represents the
effects of surfaces of zero renormalised area, $\cA  \sim O(\D^2)$; it
does not depend upon $u$ or $\mu$.
On the other hand $u\l$ represents an induced free energy per unit
length of the  polymer due to fluctuations of the finite $\cA$
surfaces; it is non-analytic in $\mu$ from eq.(\ref{string}). The
behaviour of $\l (c)$ thus governs the interesting geometrical
properties
of the polymer. In particular, fig.2 seems to show a transition
to $\l (c) =0$ at $c \sim 0.4$, which will be verified shortly.
Physically,
 as $c$ is reduced the polymer favours polymer-polymer rather than
polymer-solvent bonds and so the induced free energy $\l$ falls.
Below a certain $c_c \approx 0.4$ the polymer collapses entirely and
is
only connected into the surface by a small number of polymer-solvent
bonds. This is the coil-to-globule transition on fluctuating random
surfaces. The polymer-polymer contacts also induce free energy per
unit
length but at a lower order in $\D$, and it will be necessary to
choose
a different scaling law for $L$ in order to see this contribution
near $c = c_c$.

In order to find $c_c$ exactly, determine the universal exponent $a$,
and
the details of the phase transition, it is useful to rewrite $Z_{1}'$
as
\be
Z_{1}' = (1-c^2)^L \int_{-\pi}^{\pi} {dp \over 2\pi} \left[ \sqrt{f}
{\rm e}^{-ip} + {q \over \sqrt{f}}{\rm e}^{ip} + {t\over f^{3/2}}{\rm
e}^{3ip}
\right]^{2L} \label{pees}
\eq
As the continuum limit is approached it is expedient to rescale
$p \to p\D^d$ for some exponent $d$.
Expanding in $\D$ one has
\begin{eqnarray}
\P & =& {1\over 2c}[a_1 +a_2 ip \D^d + a_3 u  \D +a_4 p^2 \D^{2d}
+ a_5
iup \D^{d+1} + a_6 u^2 \D^2 +a_7 \mu \D^2 + \cdots] \label{long}
\end{eqnarray}
Consider first the high temperature $c \to 1$ limit.
It is known from the $c=1$ solution \cite{kostov} that $d=1/2$,  so
this ansatz will be used in the high temperature (coil) phase.
Writing $c=1-\e$ for small $\e$, the coefficients $a_i$ can be found
as power series' in $\e$; in particular $a_2 = O(\e)$ and one finds
from (\ref{pees})
\be
Z_{1}' = \sqrt{{\D}\over 4\pi l} [8 + 0(\e)]^L  {\rm
exp}\left(-ul\left(
1-{4\e \over 3} + O(\e^2)\right) \right)
\eq
in the continuum limit $\D \to 0$. This is of the form (\ref{asym}) with a $\l$
that
correctly matches onto the  answer of Fig. 2. The partition
function
itself is then
\be
Z_1   \propto N\D^{5/2}   {{\rm const}^L
\over \l l^{3/2}} \left( \sqrt{\mu} + {1\over \l l}\right)
{\rm e}^{-\l l \sqrt{\mu}} \label{form}
\eq
using eq.(\ref{string}). Large area $\cA \to \infty$ corresponds to
$\mu \to 0$, in which case the universal power exponent is $a=-5/2$.
If
$c$ is reduced by a finite amount from one, $a_2$
is finite and the resultant oscillatory
behaviour means that is no longer possible to neglect the higher orders
in $p$ as $\D \to 0$. Nevertheless one expects the form (\ref{form})
to
persist down to $c=c_c$.

Fortunately there is one other value of $c$ where $a_2$
 vanishes, given by $c=\sqrt{2} -1$; $a_3$ also
vanishes at this point indicating that this is the collapse
temperature.
A non-trivial scaling solution is obtained with the scaling laws
$c=\sqrt{2} -1 + \s \D^{1/3}$, $L= l/\D^2$, and $d=1$ (note that
polymer length scales as area now).
In this case one derives from (\ref{pees})
\be
Z_1 \propto {\rm const.}^L N\D^3 {{\rm e}^{l \mu}\over \sqrt{l}}
\int_{\mu}^{\infty}dy\ {\rm e}^{ -3 yl/2 -\a \s^3 \sqrt{y} l}
\label{comp}
\eq
in the continuum limit, where $\a \approx 41$; $a=-3/2$ at $\s
=0$.
As the reduced temperature $\s \to +\infty$ one recovers the form
(\ref{form}) with $\l \to \a \s^3$ matching in the high temperature
phase (fig.2).
At $\s=0$, $Z_1$, while now analytic in $\mu$, in this case
seems to represent {\em finite} $\cA$ surfaces. $\cA = 0$ surfaces
--- $\cA \sim O(\D^2)$ more precisely --- are characterised by the
fact
that a finite number of derivatives with respect to $\mu$ can remove
their
contribution, since each derivative brings down a factor of $\cA$.
But $Z_1(\s =0$  cannot be removed by differentiating. It can be
interpreted
as a surface dense with polymer whose area is proportional to the
polymer
length; for small $l$ differentiating does remove its
contribution.
It is tempting to extrapolate into the collapsed phase  $\s \to
-\infty$
also, but the above scaling laws do not seem to be consistent since
$Z_1$ becomes a saddle point
integral with the result $Z_1 \sim (\s^3 / l) {\rm exp}(\mu l)$. This
cannot
be correct because $Z_1$ should be monotonically decreasing with
$\mu$.

If polymers are present with fugacity $\G$ one must go back to
eq.(\ref{string}) and modify it accordingly. From eq.(\ref{two}) and
eq.(\ref{five})
one has
\be
x= -cf +{f\over c} -{3gf^2 \over c^2} + {6 gf^{3/2} \G \over c^2}
\T [i,i-1] - 2\sqrt{f} \G \T [i-1,i] \label{new}
\eq
In order that the new terms contribute to the scaling part of the
equation,
but do not disturb the critical condition for continuum
surfaces, $g_c = {1 \over 12}$ to leading order in $\D$,
one must tune $\G \to 0$ appropriately. For $c>\sqrt{2}-1$
one finds at order $\D^2$ in eq.(\ref{new})
\be
0 = \mu +z -u^2 -\g \l \sqrt{l} {\rm e}^{-l\l u} \ . \label{newcon}
\eq
The wavefunction
renormalisation
$G$ of $Z_1$ has been absorbed in $\G$ and a renormalised fugacity
$\g$
defined, of dimension $\D^{5/2}$. When $\g \neq 0$ the singularity
at $g=g_c$ ($\mu=0$)
is now shifted by an $O(\D^2)$ term, $\mu =\mu_c$ say,
which therefore couples
to renormalised area $\cA$. $\mu_c$ represents the  free energy of the
polymer system per unit $\cA$ on finite surfaces.
The positive root of eq.(\ref{newcon}) is the correct one since
$u \to +\sqrt{\mu +z}$ as $\g \to 0$. A singularity in $u$ appears when
$2u = \g \l^2 l^{3/2} {\rm e}^{-l\l u}$.
Then $u(\mu)$ is singular at $\mu=\mu_c$.
In particular at $c= \sqrt{2} -1 + \e$,
$\e >0$, one finds $u(\mu_c)  = \half \a^2 \g l^{3/2} \e^6 + O(\e^7) $ and
$\mu_c  = \a  \g \sqrt{l} \e^3 + O(\e^4)$.
For $\e <0$, $\l =0 = \mu_c$, so that $\partial^3 \mu_c / \partial c^3$
is discontinuous implying that the collapse transition is third order
The numerical factor $\a$ is  not expected to be
universal
but the $\g$ and $l$ dependence should be.
Finally, it would be interesting to study the possibility of
inter-polymer
aggregation in the low temperature phase.

\vspace{10mm}
\noindent Acknowledgements: I have benefitted from valuable
discussions
with J.Cardy, I.Kostov, and especially S.Flesia. Part of the work was
done while visiting Service de Physique Th\'eorique de Saclay and the
Aspen Center for Physics.

\begin{center}
FIGURE CAPTIONS
\end{center}
\noindent Figure 1. -- A portion of a two-dimensional simplicial
lattice made of square simplices (broken lines)
with the polymer chain shown as a
continuous line. This figure shows one contact interaction on edge
`a' (`b' is not
counted as a contact).

\noindent Figure 2. -- The induced free energy per unit length $u\l$
estimated from eqs.(\ref{sum})(\ref{derive}) at $L=150$.
Broken lines are the leading perturbative results
around $c=1$
and
$c=\sqrt{2} -1$ for $L = \infty$.
\vfil

\newpage
\begin{thebibliography}{9999}
\bibitem{coil} G.Swislow, S-T.Sun, I.Nishio, and T.Tanaka, Phys. Rev.
Lett.
{\bf 44} (1980) 796.
\bibitem{dg} P.G. de Gennes, {\em Scaling concepts in Polymer
Physics},
Cornell University Press, Ithaca, NY, (1979).
\bibitem{jan} J.Ambjorn, B.Durhuus, and J.Frohlich, \NP {\bf B257}
(1985) 433;
\bibitem{rest} F.David, \NP {\bf B257} 45 (1985);
V.A.Kazakov, I.K.Kostov, and A.A.Migdal,
\PL {\bf B157} (1985) 295.
\bibitem{dup} B.Duplantier and H.Saleur, \PRL {\bf 59} (1987) 539;
\NP {\bf B290} (1987) 291;
D.Dhar and J.Vannimenus, J. Phys. {\bf A20} (1987) 199.
\bibitem{kostov} B.Duplantier and I.Kostov, \NP {\bf B340} (1990) 491.
\bibitem{mehta} D.Bessis, C.Itzykson, and J-B.Zuber, Adv. in Appl.
Math.
{\bf 1} (1980) 109; M.L.Mehta, Commun. Math. Phys. {\bf 79} (1981)
327.
\end{thebibliography}
\end{document}